# A Unified Anti-Jamming Design in Complex Environments Based on Cross-Modal Fusion and Intelligent Decision-Making

Huake Wang*, *Member*, *IEEE*, Xudong Han, Bairui Cai,

Guisheng Liao, *Senior Member*, *IEEE*, and Yinghui Quan, *Senior Member*, *IEEE*

*Abstract*—With the rapid development of radar jamming systems, especially digital radio frequency memory (DRFM), the electromagnetic environment is increasingly complicated. In recent years, neural networks have been successfully applied in the fields of radar interference recognition and anti-jamming, such as convolutional neural networks (CNNs). However, most existing studies focus solely on either jamming recognition or anti-jamming strategy design. In this paper, we propose a unified framework that integrates interference recognition with intelligent anti-jamming strategy selection. Specifically, time-frequency (TF) features of radar echoes are first extracted using both Short-Time Fourier Transform (STFT) and Smoothed Pseudo Wigner–Ville Distribution (SPWVD). A feature fusion method is then designed to effectively combine these two types of time-frequency representations. The fused TF features are further combined with time domain features of the radar echoes through a cross-modal fusion module based on an attention mechanism. Then, a three-class classification algorithm is employed to recognize different interference types. Finally, the recognition results, together with information obtained from the passive radar, are fed into a Deep Q-Network (DQN)-based intelligent anti-jamming strategy network to select jamming suppression waveforms. The key jamming parameters obtained by the passive radar provide essential information for intelligent decision-making, enabling the generation of more effective strategies tailored to specific jamming types. The designed method demonstrates improvements in both jamming type recognition accuracy and the stability of anti-jamming strategy selection under complex environments. Experimental results show that the presented method gets superior performance compared to Support Vector Machines (SVM), VGG-16, 2d-CNN methods, with respective improvements of 1.41%, 2.5%, and 14.51% in overall accuracy. Moreover, in comparison with the SARSA algorithm, the designed algorithm achieves faster reward convergence and more stable strategy generation.

This work was supported in part by the National Natural Science Foundation of China under Grants 62301410 and 62331019, in part by the National Key Laboratory of Air-based Information Perception and Fusion under Grants ZH 2024-0053.
Huake Wang and Yinghui Quan are with School of Information Mechanics and Sensing Engineering, Xidian University, Xi'an, Shaanxi, China, 710071, (e-mail: hkwang@xidian.edu.cn; yhquan@mail.xidian.edu.cn).
Huake Wang, Xudong Han, Bairui Cai, Guisheng Liao and Yinghui Quan are with Hangzhou Institute of Technology, Xidian University, Hangzhou, Zhejiang, China, 311231, (e-mail: hkwang@xidian.edu.cn; xudhan@stu.xidian.edu.cn; youngbarry@126.com; gsliao@xidian.edu.cn).

## I. INTRODUCTION

RADAR systems have the operating capability in long distance by transmitting high-frequency electromagnetic waves and analyzing the echoes, also can work long time in all-weather. It plays a vital role in modern warfare by providing crucial situational awareness, such as target tracking, velocity estimation, and imaging. However, with the rapid development of radar jamming systems, especially digital radio frequency memory (DRFM), the electromagnetic environment is increasingly complicated [1]. Because of the increasing forms of jamming have emerged, there is a need to discern the jamming accurately. The active jamming can be generally divided into two categories: suppression jamming and deception jamming [5]. Suppression jamming typically employs noise signals to cover the true target echoes, thereby reducing the signal-to-noise ratio (SNR) and interfering with the radar system, for example the Aiming Jamming (AJ). Deception jamming often utilizes linear frequency modulation (LFM) signals and retransmits signals real echo-like, such as Range False Target Deception Jamming (RFTJ), and Range Dense False Target Jamming (RDFTJ). It has high-power at different time instances to create false targets, misleading the radar system. Smart noise jamming is capable of producing both deceptive effects and suppression functions at the same time, making it a hybrid and more challenging form of jamming. In order for the radar to function properly, the corresponding jamming must be separated or suppressed. Therefore, the jamming recognition plays a decisive role in anti-jamming technologies, as it directly influences the selection of appropriate countermeasures.

In recent years, artificial intelligence (AI) has been widely applied in radar systems to enhance anti-jamming capabilities [7]. In particular, AI-based models have demonstrate strong performance in tasks such as jamming type recognition and intelligent waveform decision-making for frequency-agile radar. However, most existing research focuses on these two aspects independently. There is still a lack of integrated frameworks that can simultaneously identify the jamming type and adaptively select the appropriate anti-jamming waveform, which is essential for achieving robust and intelligent radar operation in dynamic electronic warfare environments.

In jamming type recognition work, with continuous exploration, Convolutional Neural Network (CNN), as one of the core deep learning methods, have been successfully applied in radar jamming type recognition due to their superior feature learning capabilities. Many studies have focused on optimizing the CNN architecture or processing the input signal data to improve recognition performance. In [10], the authors designed a CNN-based method that performs radar jamming type recognition using only time domain features. To enable more complex jamming recognition, researchers have started to employ deeper-level features as input. The authors proposed using CNNs to extract spatial features of jamming signals for recognizing radar compound jamming signals [11], and extracting the time-frequency (TF) images of jamming signals by short-time Fourier transform (STFT). After that these images are simply cut to eliminate redundancy, and then put into CNN for recognition was proposed [13]. However, when faced with limited training data, the above methods often fail to achieve satisfactory performance. To improve performance under limited training data conditions, an applied data augmentation in CNN-based network [15] was established, a method using multi-scale image features are extracted using ResNet-50 and a Feature Pyramid Network (FPN), which is fused as a unified representation [16]. To further enhance recognition performance, researchers have been continuously refining network architectures and methodologies. In [17], a weighted ensemble CNN with transfer learning was proposed to enhance model stability and recognition accuracy. An adversarial deep learning was employed to tackle jamming attacks and defense strategies in cognitive radio security [18]. In [19] the authors proposed a deep learning network-based open set radar jamming recognition method. But the compound features of multi-type jamming signals have not been fully addressed. handling complex jamming patterns proposed AC-VAEGAN, combining Generative Adversarial Networks and Variational Autoencoders [20]. In [21] the authors proposed a method based on spectrum waterfall graphs, which is effective for training on radar signal spectrograms.

DRL has demonstrated significant potential in radar anti-jamming decision-making due to its adaptability and strategy optimization capabilities. In [22], each radar independently learns a frequency hopping strategy through a local Q-table to avoid the frequency bands occupied by jammers and other radars, while employing an ε-greedy policy to balance exploration and exploitation. Authors integrated transfer learning into the DRL framework, facilitating knowledge transfer between similar tasks and accelerating the learning process [23]. Researcher proposed a dual-layer decision-making architecture combining Deep Deterministic Policy Gradient (DDPG) and Multi-Agent DDPG to address high-dimensional action spaces and complex jamming environments [24]. In [25] the authors signed a DRL-based anti-jamming strategy for frequency-agile radar, enhancing resistance to frequency domain jamming boosting the efficiency of intelligent strategy decisions.

Meanwhile, researchers have proposed using cooperative radar systems to improve the overall performance of radar systems. in [26] the authors utilized a Deep Q-Network (DQN) algorithm, combined with radar's sensing capabilities, to identify and respond to jamming signals. These using cooperative radar systems allow radars to share resources and complement each other's functions showing better measurement accuracy than single radars [29]. For deceptive jamming with false targets, three discrimination methods based on azimuth, radial distance, and 2D angle have been proposed using active-passive heterogeneous radar systems. All three methods can effectively identify false targets while maintaining high recognition accuracy for true targets [31].

To deal with the issue of integrated design for jamming type recognition and subsequent anti-jamming strategy during anti-jamming operation. An intelligent anti-jamming network is proposed in this paper to first recognize jamming type and then use the output to generate anti-jamming waveforms. The main contributions and novelties of our work are as follows:
1) We designed a H-ResNet based on deep CNN to extract more abstract features from the TF domain. After that we proposed a cross-modal fusion network to fusion the TF domain features and time domain features. In addition, a cross-modal fusion network was developed to integrate the both domain features. These methods can capture more comprehensive signal characteristics.
2) We customized reward functions for AJ, RFTJ, and DRFTJ within the intelligent waveform decision-making network to address different jamming scenarios. These reward functions guide the agent with clear objectives and prevent it from learning irrelevant behaviors. This method can facilitate it escaping from suboptimal policies during training.
3) We propose an integrated anti-jamming framework that combines jamming type information from active radar and parameter data from passive radar. The fused data is processed by an intelligent strategy network, which enhancing interference resistance by adaptively adjusting waveform parameters based on the jamming environment.

The remaining of this paper is organized as follows. Section II presents a related work of fundamental methods. Section III describes the proposed method in detail. Section IV evaluates the performance of the proposed method experimentally. Finally, Section V concludes this article.

## II. RELATED WORK

This section introduces the radar signals, different types of jamming, principles of TF analysis methods, CNN, and DQN.

### A. LFM Signal

The radar transmits a LFM signal $s_t(t)$ [33],

$$s_t(t) = rect(\frac{t}{T}) \cdot exp(j(2\pi f_0 t + \pi \mu t^2)) \tag{1}$$

where $rect(\frac{t}{T})$ is a rectangular function with time width $T$, $f_0$ is the center frequency of the carrier, $\mu = \frac{B}{T}$ is the frequency modulation slope of LFM, $B$ is bandwidth.

## B. Jamming Signals

Radars rely on the analysis of echo observing objects to obtain relevant parameters. In radar system, the Signal-to-Interference-plus-Noise Ratio (SINR) is a fundamental metric governing detection performance. A higher SINR makes it easier to detect targets. The jamming signals always to minimize the received SINR. It can jam and obstruct radar systems to detect objects normally. In this work, three types of jamming scenarios are constructed as followed.

1) *AJ:* It is a type of intentional jamming where the jammer attempts to match the carrier frequency of the radar signal in real time, transmitting wideband or narrowband noise to degrade radar performance.

$$J_1(t) = n(t) \cdot rect(\frac{t}{T_J}) \cdot \cos(2\pi f_J t + \varphi_J) \quad (2)$$

where $n(t)$ the Additive White Gaussian Noise with zero mean and variance, $T_J$ is the duration of jamming, $f_J$ is the carrier frequency of jammer, and $\varphi_J$ is a random initial phase.

2) *RFTJ:* It involves transmitting delayed replicas of the radar's signal to create false targets. This type of jamming aims to mislead the radar's range estimation by inserting deceptive echoes at incorrect distances.

$$J_2(t) = s_t(t - \Delta\tau) \cdot e^{j2\pi\Delta f t} \quad (3)$$

where $\Delta\tau$ is the transmit delay, and $\Delta f$ is the doppler frequency shift.

3) *RDFTJ:* It is an advanced electronic attack technique. The jammer captures radar's transmitted signal then retransmits multiple delayed and modified copies to generate a cluster of false targets. These false returns can obscure the true target or overwhelm the radar's signal processing system.

$$J_3(t) = \sum_{k=1}^{K} s_t(t - \Delta\tau_k) \cdot e^{j2\pi\Delta f_k t} \quad (4)$$

where $\Delta\tau_k$ is the transmit delay of the $k-th$ false target, and $\Delta f_k$ is the doppler frequency shift of the $k-th$ false target.

## C. TF Analysis Methods

Compared with single frequency domain features, TF features [13] inherently capture the joint energy distribution of signals across both time and frequency domains, which providing a more comprehensive characterization of features.

1) *STFT:* The STFT [34] enables time-localized spectral analysis of non-stationary signals by partitioning the signal into windowed segments. It can capture time-varying frequency components. It divides the signal into multiple windowed segments in the time domain, then performs Fourier transform on each segment, defined by:

$$X(\omega,t) = \int_{-\infty}^{\infty} x(\tau)h(\tau-t)e^{-j\omega\tau} d\tau \quad (5)$$

where $x(\tau)$ is the original time domain signal, $h(\tau-t)$ denotes the analysis window centered at time $t$.

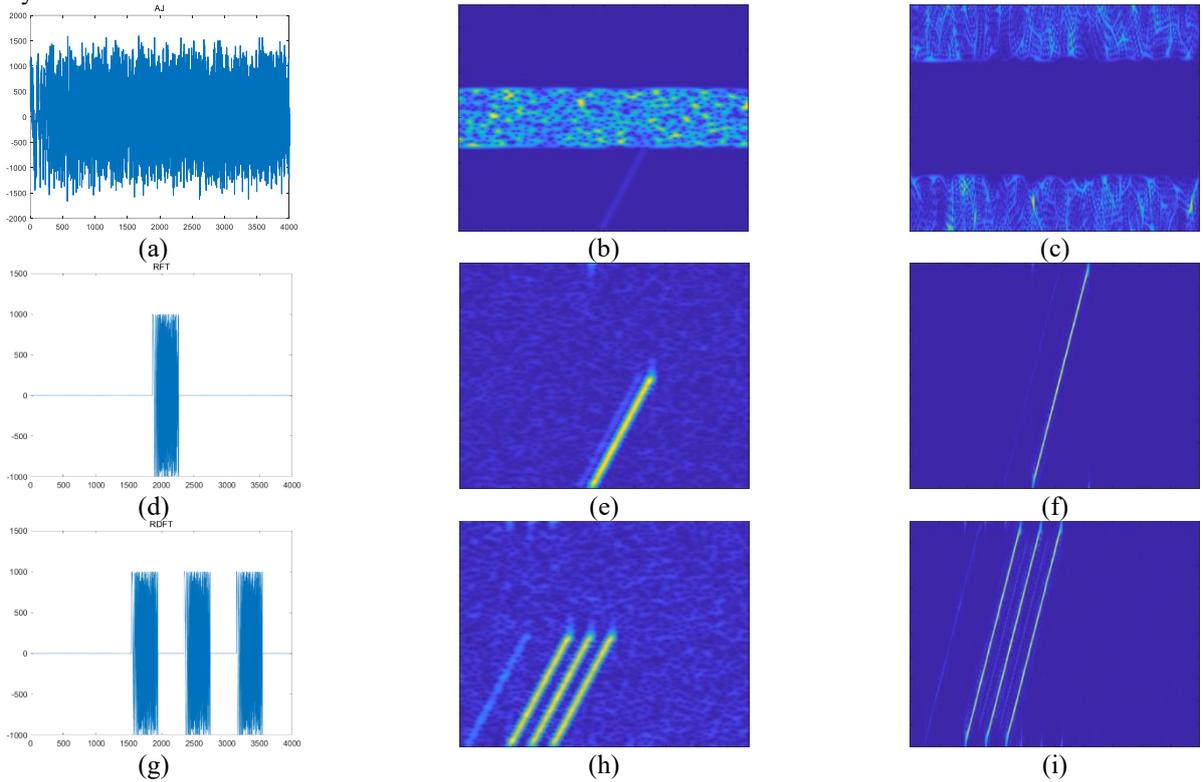

Fig. 1. Time domain waveforms and time-frequency spectrograms of the three interference types. (a) AJ. (b) AJ-STFT. (c) AJ-SPWVD. (d) RFT. (e) RFT-STFT. (f) RFT-SPWVD. (g) RDFT. (h) RDFT-STFT. (i) RDFT-SPWVD.

2) *Smoothed Pseudo Wigner-Ville Distribution (SPWVD):* Unlike the STFT, which suffering from fixed resolution limitations due to windowing. The SPWVD [35] employs independent smoothing operations in time and frequency. It enables flexible trade-offs between auto-term concentration and cross-term artifacts—a critical advantage for analyzing rapidly varying signals. It is evolved from Wigner-Ville Distribution, it applies dual smoothing kernels

to suppress cross-term interference while retaining high TF resolution. This operation is expressed as:

$$S(t,\omega) = \int x(t-v+T/2) \cdot x^*(t-v-T/2) \cdot h(T)g(v)e^{-j\omega\tau}dvdT \quad (6)$$

where $x(t)$ is analytic signal, $h(T)$ and $g(v)$ is window function which $h(T)$ is frequency domain smoothing window, $g(v)$ is time domain smoothing window.

*D. DQN*

DQN [26] is a value-based reinforcement learning algorithm that integrates Q-learning with deep neural networks to estimate the action-value function. It enables agents to make decisions in complex and high-dimensional environments by approximating Q-values directly from raw input features. To enhance training stability, DQN incorporates experience replay, which breaking correlation between sequential data, and reducing oscillations during learning. This approach has demonstrated strong performance in a variety of sequential decision-making tasks.

## III. PROPOSED APPROACH

*A. Recognition Network Design*

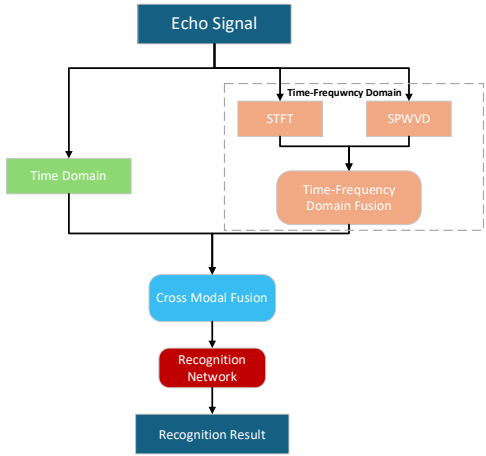

Fig. 2. Flowchart of the jamming recognition

through the aforementioned two methods, the resulting TF images are resized $224 \times 224$ to uniform dimensions and normalized. To fully exploit the information between these different TF modalities, a hierarchical ResNet (H-ResNet) structure [36] is employed for deep feature extraction. Then the two feature vectors from different branches are concatenated to form a unified representation.

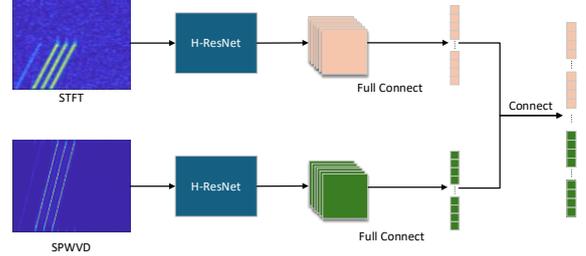

Fig. 3. TF Feature extraction and fusion process

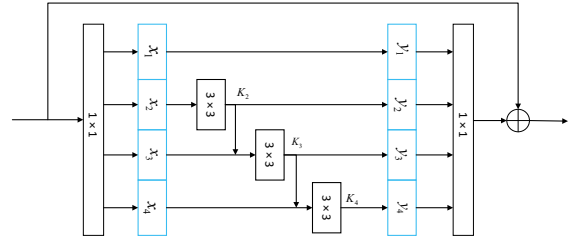

Fig. 4. H-ResNet module

2) Cross-Modal Fusion Network: In our cross-modal fusion model, the features involved in the fusion process including time domain features and the aforementioned fused features. First the temporal features are processed through $7 \times 7$ convolutional layer and $3 \times 3$ max-pooling layer. Then they pass through a residual block composed of $1 \times 1$ and $3 \times 3$ convolutional layers. The $1 \times 1$ convolutional kernels perform feature vector dimension expansion and reduction. Subsequently, these features are passed through an average pooling layer and a fully connected layer [37]. Finally, classification of the three jamming signals is accomplished by a linear layer with *Softmax* activation, while using a cross-entropy loss function to optimize the regression work.

1) TF Domain Features Fusion Network: In this fusion network, extracting TF representations of radar signals

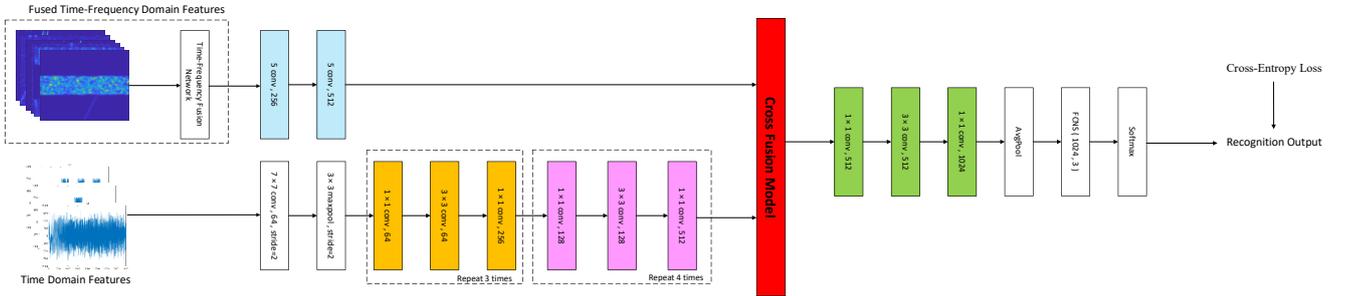

Fig. 5. Overall flowchart of the proposed jamming recognition algorithm.

After processing through convolutional layers, the input data are decomposed into multiple feature vectors. Subsequent residual mapping operations convert these features into query vectors $Q = \{q_i\}$, key vectors $K = \{k_i\}$ and value vectors $V = \{v_i\}$ for attention computation. Let

$F_{t/tf}$ denote the feature vector after the time domain and TF domain features making cross-modal processing, where $t$ is corresponding time domain branching, $tf$ is TF domain branching. These $F_{t/tf}$ convert into $Q_{t/tf}$, $K_{t/tf}$ and $V_{t/tf}$.

Then the self-attention mechanism's feature vector $A_{t/tf}$ is computed using the following function:

$$A_{t/tf} = MHA(Q_t, K_{tf}, V_{tf}) \quad (7)$$

Global Average Pooling is then applied to $A_{t/tf}$ for dimensionality reduction to prevent overfitting. A *Sigmoid* activation function normalizes channel values to the range $[0,1]$, where values closer to 1 indicate higher channel importance. These normalized values serve as influence weights. The original features are weighted through matrix multiplication with the influence weights, as expressed by:

$$\tilde{F}_{t/tf} = \sigma(GAP(A_{t/tf})) \odot F_{t/tf} \quad (8)$$

the symbol $\odot$ denotes the *Hadamard product*, where each element in the resulting matrix is the product of the corresponding elements in the input matrices.

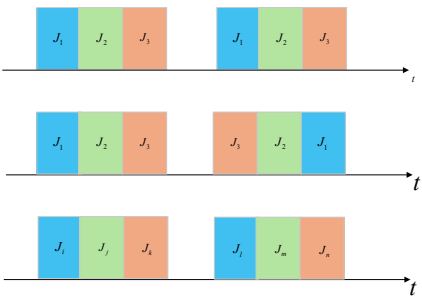

Fig. 6. Cross fusion module

In this work, the classifier is implemented with a multilayer perception (MLP) architecture [40], employing *ReLU* activation functions and *dropout* layers for overfitting mitigation. The *ReLU* activation functions expressed as:

$$\begin{cases} F(x) = max(0, x) \\ x = \tilde{F}_{t/tf} \end{cases} \quad (9)$$

*B. Intelligent Waveform Decision Network*

1) Jamming Policy Design: In jamming generation strategy, this paper systematically coordinates AJ, RFTJ and RDFTJ according to specific scheduling policies.

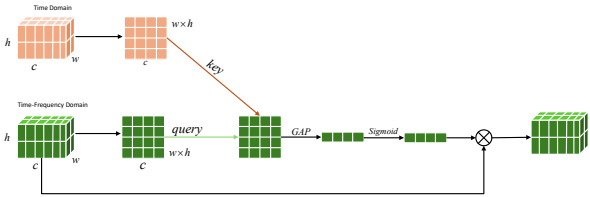

Fig. 7. Layout of jamming deployment

Jamming type J1 represents AJ, J2 represents RFTJ, and J3 represents RDFTJ. Jamming strategy Ⅰ indicates sequential switching among the three jamming types, with one type switched per CPI. Strategy Ⅱ denotes palindromic-sequence switching of the three jamming types, while strategy Ⅲ represents random switching, where the system randomly transitions to another jamming type after each CPI.

2) Radar Anti-Jamming Waveform Decision Network: Deep reinforcement learning policy network evaluates the effectiveness of anti-jamming decisions in the previous time step through a reward function computation. It input the state vectors and rewards into action-decision network in the form of $(s_t, r_t, s_{t+1})$, which $s_t$ is the state vector at time $t$, $r_t$ is the reward vector at time $t$, and $s_{t+1}$ is the state vector at time $t+1$.

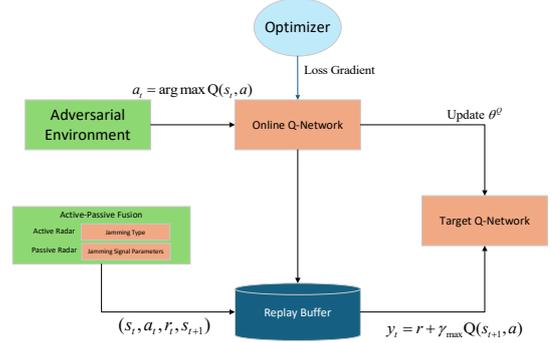

Fig. 8. Framework of the anti-jamming decision-making strategy

This Q-network model employs dual network: an online Q-network and target Q-network, following the fixed target network mechanism proposed in the original DQN framework [22] [25]. The two networks share the same architecture but maintain separate parameters.

In the Q-network architecture, the input is the current state vector, and the output is the Q-value corresponding to each possible discrete action. The target Q-network has the same structure as the online Q-network but is updated at regular intervals to stabilize learning.

The online Q-network is trained to minimize the temporal difference (TD) error between the predicted Q-value and the target Q-value. The target Q-value is computed using the Bellman equation:

$$y_t = r + \gamma_{max} Q(s_{t+1}, a) \quad (10)$$

where Q is the target Q-network. The mean squared error between $y_t$ and $Q(s_t, a_t)$ is used as the loss function for gradient descent optimization.

An experience replay buffer is used to store transitions $(s_t, a_t, r_t, s_{t+1})$, and training is performed using mini-batches sampled randomly from this buffer. This improves sample efficiency and helps break the correlation between consecutive samples, thus enhancing the stability of the training process.

3) Reward Function Design: In anti-jamming scenarios, a well-designed reward function can significantly enhance the convergence speed and optimal performance of reinforcement learning algorithms. This technique evaluates the radar's anti-jamming decision-making performance mainly from short-term behavior assessment.

The short-term behavior evaluation framework is characterized by two indicators: the SINR obtained by the radar as well as the feature relationship between the jammer and radar signals, as expressed by:

$$SINR = \frac{P_s h_s^2 \sigma}{P_n + P_j h_s I(f_j = f_n)} \quad (11)$$

where $P_s$ is the radar transmit power, $P_n$ is the power of the environmental noise received by the radar, $P_j$ denotes the power of the jammer, $h_s$ denotes the channel gain from the radar to the target jammer, $f_n$ is the carrier frequency of radar's $n-th$ pulse, $f_j$ is the frequency of the jammer.

If $f_j = f_n$, then $I = 1$; otherwise, $I = 0$. A larger SINR indicates that the target signal is more easily detected by the receiver, leading to better performance of the radar system.

In the short-term behavior evaluation framework, specific reward functions are designed for three types of jamming. The need for active-passive radar fusion is very important in this work, which enabling higher-accuracy jamming signal sorting. For AJ, the radar adopts a frequency-agile anti-jamming waveform, and the single-step reward function is expressed as follows [41]:

$$R = \begin{cases} 30, & \text{if } signal_{high} < jam_{low} \text{ or } signal_{low} > jam_{high} \\ SINR, & \text{if } signal_{low} < jam_{low} < signal_{high} \\ & \text{or } signal_{low} < jam_{high} < signal_{high} \\ -100 & \text{if } jam_{low} < signal_{low} < signal_{high} < jam_{high} \end{cases}$$
(12)

when the radar is not interfered with, the reward is set to 30. If the radar is partially interfered with by noise signals, the reward is represented by the SINR. In the case of complete jamming, the reward is set to –100.

For RFTJ, the radar employs a frequency-agile anti-jamming waveform too. The action space remains the same as that used for noise-directed jamming. However, to ensure orthogonality among the transmitted frequencies in the anti-deception waveform, the frequency hops must be integer multiples of a base spacing $\Delta f$. Considering the cost associated with frequency hopping, a cost value $c$ is introduced to quantify the trade-off between performance and resource consumption. Therefore, the single-step reward function is expressed as follows:

$$R = \begin{cases} 30 - c \cdot \max(\frac{|f_{S+1} - f_S|}{\Delta f}, b), & \text{if } |f_{S+1} - f_S| = n\Delta f, \\ & \text{where n is a positive integer} \\ SINR & , \text{if } f_{S+1} = f_S \\ -\max(n) & , \text{else} \end{cases}$$
(13)

where $f_{signal}$ is the frequency of the transmitted waveform, $\Delta f$ is the minimum frequency spacing, $c$ is the frequency-hopping cost and $b$ is the allowable range of the frequency-hopping cost.

TABLE I  RADAR JAMMING INFORMATION

| Jamming type | Parameters | Range of values |
|---|---|---|
| AJ | Carrier Frequency | 1GHz |
| | Transmit Bandwidth | 28MHz |
| | Interference Bandwidth | 60MHz |
| | JNR | 10~35dB |
| | Angle of the jammer | 2° |
| RFTJ | Carrier Frequency | 1GHz |
| | Transmit Bandwidth | 28MHz |
| | Target Delay | 10~20 $\mu s$ |
| | JNR | 10~35dB |
| | Angle of the jammer | 2° |
| RDFTJ | Carrier Frequency | 1GHz |
| | Transmit Bandwidth | 28MHz |
| | Number of false targets | 3~5 |
| | Target Delay | -15~15 $\mu s$ |
| | JNR | 10~35dB |
| | Angle of the jammer | 2° |

For RDFTJ, the radar typically employs a cover-pulse anti-jamming waveform. The single-step game reward function is expressed as follows:

$$R = \begin{cases} 30, & \text{if } t_{cheat} > t_{observe} \\ SINR, & \text{else} \end{cases}$$
(14)

where $t_{cheat}$ represents the time at which the radar transmits the cover pulse, and $t_{observe}$ represents the observation window duration of the jammer.

4) Active-Passive Radar Fusion: When relying solely on active radar, the narrowband channel exhibits limited accuracy in identifying the carrier frequency and bandwidth of jamming signals. In contrast, the wideband channel of passive radar provides more precise characterization of these parameters.

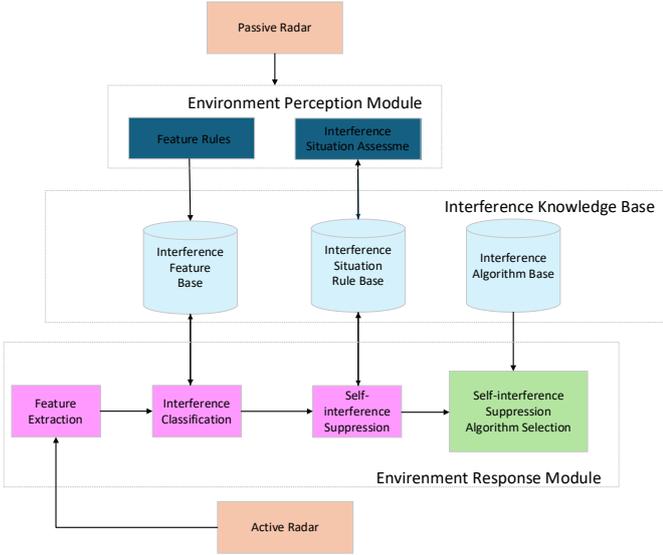

Fig. 9. Framework of the Active–Passive radar cooperative framework

## IV. EXPERIMENTAL RESULTS

In this section, first we provided a detailed overview of the jamming dataset, followed by a description of the experimental settings for the proposed Fusion-based Deep Jamming Recognition (FDJR) and the comparison algorithms. Subsequently, a set of evaluation metrics is employed to report the experimental results and perform a comparative analysis.

### A. Dataset Generation

To evaluate the recognition performance of the proposed recognition network, this work simulates the three jamming signals introduced in Section Ⅱ, whose detailed parameters could be found in Table Ⅰ. The LFM waveform is used as the radar transmission signal, where the transmitted signal is configured with a carrier frequency of $1\,\text{GHz}$ and a transmission bandwidth of $28\,MHz$. The pulse duration is $10\,\mu s$, and the pulse repetition interval is $50\,\mu s$. SNR is set to $10\,dB$. An antenna array with 8 elements is employed, and the target is positioned at an angle of $5°$.

### B. Experiment Settings

The backbone network structure of features extraction network designed in this paper is shown in Table Ⅱ and Table Ⅲ.

TABLE IV TIME DOMAIN FEATURE EXTRACTION

| Layer | Conv | Repetition | Activation | Dropout |
|---|---|---|---|---|
| Input | | $224\times 224$ | | |
| Conv.1 | $7\times 7$ | 1 | ReLu | 0.4 |
| MaxPool | $3\times 3$ | 1 | - | - |
| Conv.2 | $3\times 3$ | 5 | ReLu | 0.5 |
| Conv.3 | $1\times 1$ | 1 | - | - |

TABLE V TF FEATURE EXTRACTION

| Layer | Conv | Repetition | Activation | Dropout |
|---|---|---|---|---|
| Input | | $224\times 224$ | | |
| Conv.1 | $7\times 7$ | 1 | - | - |
| MaxPool | $3\times 3$ | 1 | - | - |
| Conv.2 | $1\times 1$ | 3 | - | - |
| Conv.3 | $3\times 3$ | 3 | ReLu | 0.5 |
| Conv.4 | $1\times 1$ | 3 | - | - |

The repetition count of 3 indicates that a group of Conv.2, Conv.3, Conv.4 layers is used as a unit, and this unit is repeated three times in succession.

To validate the advantages of proposed algorithm in jamming type recognition, we employed different methods to extract features to fed them into baseline networks, including VGG, SVM, and CNN. The proportion of datasets used for model training, testing and validation is 0.3:0.6:0.1.

The experiments all were conducted on a workstation equipped with 64 GB RAM, an Intel i7-14700K CPU, and an NVIDIA RTX 3090Ti GPU. The software environment included Python 3.9.6, PyTorch 2.4.1, and CUDA 12.1.

### C. Evaluation Metrics

In recognition model, there are four fundamental metrics to calculate other metrics: True Positive (TP), True Negative (TN), False Positive (FP), and False Negative (FN), the three-class classification model produces decision label vectors $\hat{y}_i$.

To verify the effectiveness of the jamming type recognition algorithm, this article analyzes from the following evaluation metrics.

1) *Overall Accuracy (OA):* OA is the number of correctly predicted samples $N_{\text{correct}}$ refers to the number of instances where the predicted label matches the true label. OA is computed by dividing this number by the total number of samples, and its expression is given as follows:

$$OA = \frac{N_{\text{correct}}}{N_{total}} \times 100\% \quad (15)$$

2) *Recall:* Recall is employed to evaluate the precision of the model in identifying positive samples of the jamming class. It can be computed for each jamming type. For the $j-th$ jamming, the recall is defined as follows:

$$Recall_j = \frac{TP_j}{TP_j + FN_j} \quad (16)$$

3) *Precision:* Precision reflects the accuracy of the model in predicting positive examples for each jamming class. It measures the proportion of correctly predicted positive samples out of all samples predicted as positive. For the $j-th$ jamming type, precision is defined as:

$$Precision_j = \frac{TP_j}{TP_j + FP_j} \quad (17)$$

4) *F1-Score (F1):* F1 provides a harmonic mean of precision and recall, offering a balanced evaluation metric especially in cases of class imbalance. For the $j-th$ jamming type, the F1 is given by:

$$F1_j = \frac{2 Precision_j \cdot Recall_j}{Precision_j + Recall_j} \quad (18)$$

In decision-making network based on deep reinforcement learning to evaluate the performance we use average reward and convergence speed.

5) *Convergence Speed (CS):* CS quantifies the rate at which a model minimizes its loss function and approaches optimal performance during the training process. Faster convergence indicates that the model is learning efficiently. This is especially useful when computational resources or time are limited. It also reflects the effectiveness of the network architecture design, the optimization algorithms, and the data preprocessing strategies. Therefore, it serves as a practical metric for assessing training efficiency and algorithm scalability.

6) *Stability:* Stability refers to the consistency and smoothness of a model's learning behavior during the optimization process. A stable model should exhibit gradual and monotonic improvement in its performance metrics over time, without sharp fluctuations or divergence. It provides insight into the reliability of the training process and the model's robustness to initialization or data-related variations.

### D. Simulation Experimental Results and Analysis

In this part, Table VI presents per class OA, Recall, Precision, and F1-score of the SVM [43], VGG16 [44], 2D-CNN [45], as well as the proposed algorithm with 80% training samples. The best results for each data are highlighted in bold. The experimental results in Table 4 show that the proposed deep learning network is butter than the compared four algorithms.

TABLE VII   JR RESULTS, RESPECTIVELY, OBTAINED BY SVM, VGG16, 2D-CNN, FDJR

| Training:80% | SVM | VGG16 | 2d-CNN | FDJR |
| --- | --- | --- | --- | --- |
| OA (%) | 91.70 | 93.69 | 93.13 | 95.45 |
| Recall (%) | 87.08 | 89.59 | 88.49 | 93.43 |
| Precision (%) | 93.48 | 96.18 | 95.09 | 96.34 |
| F1-score (%) | 89.52 | 92.17 | 91.11 | 94.71 |

The overall recognition performance at different proportion of training set of five methods is shown in TABLE VIII for straight and convenient comparisons. Specifically, at the proportion of training is 80%, the proposed algorithm achieves improvements of *3.75%, 1.76%, 2.32%* in OA, *6.35%, 3.84%, 4.94%* in Recall, *2.86%, 0.16%, 1.25%* in Precision, and *5.19%, 2.54%, 3.60%* in F1-Score compared to SVM, VGG16, and 2d-CNN. Moreover, when the training set size is limited, the proposed method consistently demonstrates superior performance across all mentioned metrics, further highlighting its robustness and effectiveness under small training datasets conditions. The confusion matrix of proposed recognition method is shown as Fig. 10

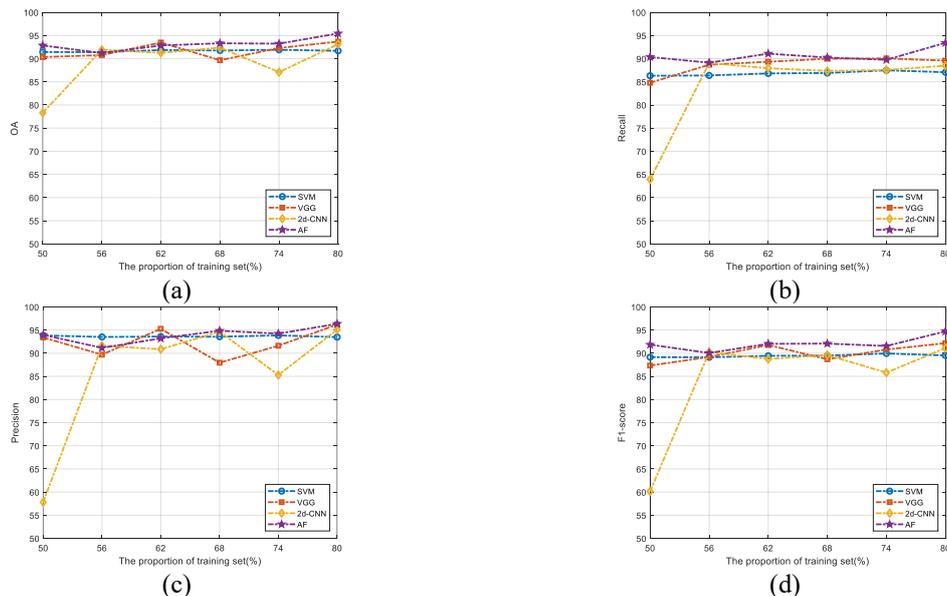

Fig. 10. Confusion matrix of the proposed jamming recognition method

Fig. 11. Evolution of the evaluation metrics according to the training set size. (a) OA. (b) Recall. (c) Precision. (d) F1.

The jamming types prediction results obtained by the designed recognition network are fed into the parameter input strategy network of the passive radar. Different jamming suppression measures are then taken based on the agent's received rewards, which evaluating the effectiveness against various types of jamming signals, as illustrated in Fig. 12. We use the SARSA algorithm to make comparative experiments and evaluate the proposed decision-making algorithm by observing the stability of the two algorithms in obtaining reward.

The trained network is evaluated under various jamming conditions. For the AJ scenario, the jammer generates targeted noise interference centered at the radar's central frequency from the previous time step, with a bandwidth 2–4 times wider than the radar signal. In the RFTJ scenario, the jammer synthesizes false targets by replicating the radar signal's frequency from the previous time step, aligning its center frequency with that of the radar, and continues until the radar performs a frequency hop to suppress the interference. In the DRFTJ scenario, the jammer similarly generates false targets based on the radar's last-used frequency, persisting until the radar executes a frequency hopping strategy. To counteract these dense false targets, the radar transmits a deceptive probing signal longer than the jammer's sensing window and then emits a frequency-hopped waveform in the subsequent time step. The decision-making behavior of the deep reinforcement learning network under these jamming conditions is illustrated in Fig. 13.

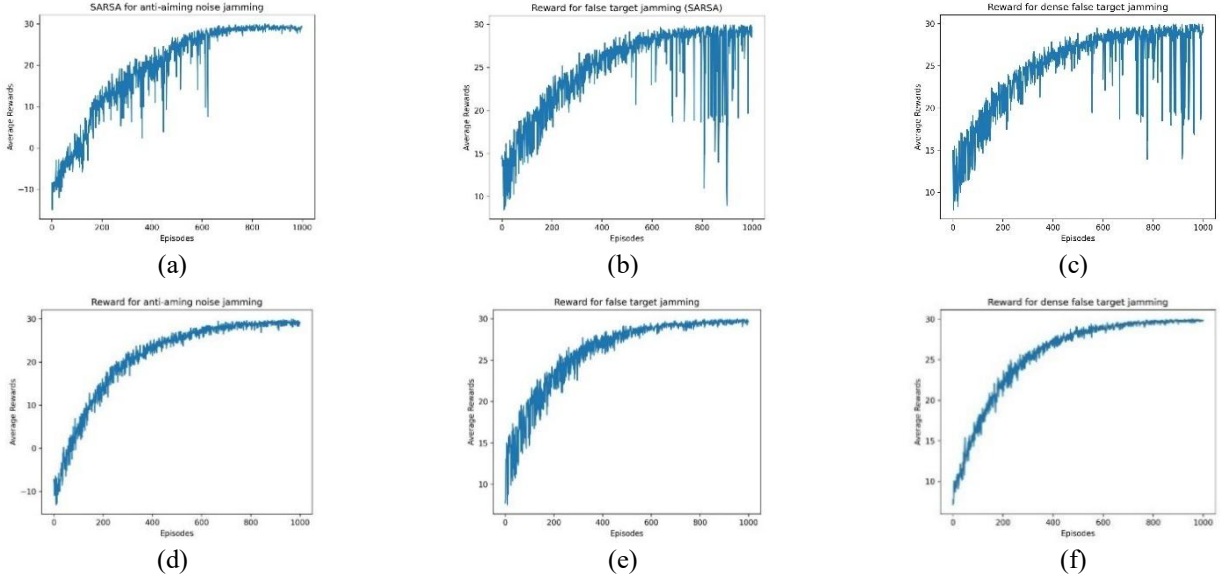

Fig. 12. Average Rewards of the SARSA and proposed algorithm. (a) AJ-SARSA. (b) RFT-SARSA. (c) RDFT-SARSA. (a) AJ-Proposed algorithm. (b) RFT-Proposed algorithm. (c) RDFT-Proposed algorithm.

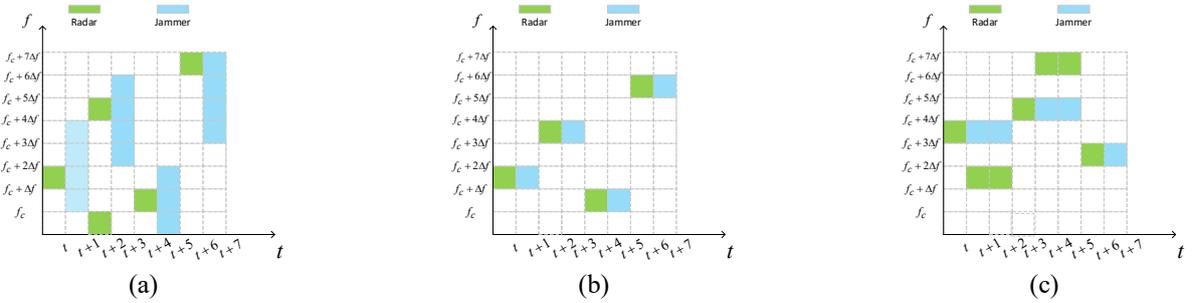

Fig. 13. Policy output of the anti-jamming strategy design network (a) AJ. (b) RFT. (c) RDFT.

As observed from the reward function curves, in SARSA algorithm, convergence under AJ is not reached until approximately episode 800, and the training phase exhibits significant oscillations in the early stages. Moreover, for RFT and RDFT, the reward curves fail to converge and instead show considerable instability during the later stages of training. The proposed method also achieves convergence around episode 600 when dealing these three jamming signals, with minimal fluctuations throughout the training process. The result indicates that the proposed method demonstrates superior robustness and stability compared to the baseline.

As shown in Fig. 14, the vertical axis encodes both the jammer type and the corresponding decision parameters of the radar and jammer. The integer part of the vertical axis values (i.e., 0, 1, and 2) represents three different jamming types. The decimal part encodes the normalized anti-jamming waveform behavior for each jamming type.

For example, a value of 0.5 indicates that the jammer employs jamming type 0 (AJ) with a normalized interference behavior. To avoid ambiguity where the maximum normalized value of 1.0 would change the integer part, the normalized behavior is divided by 2.

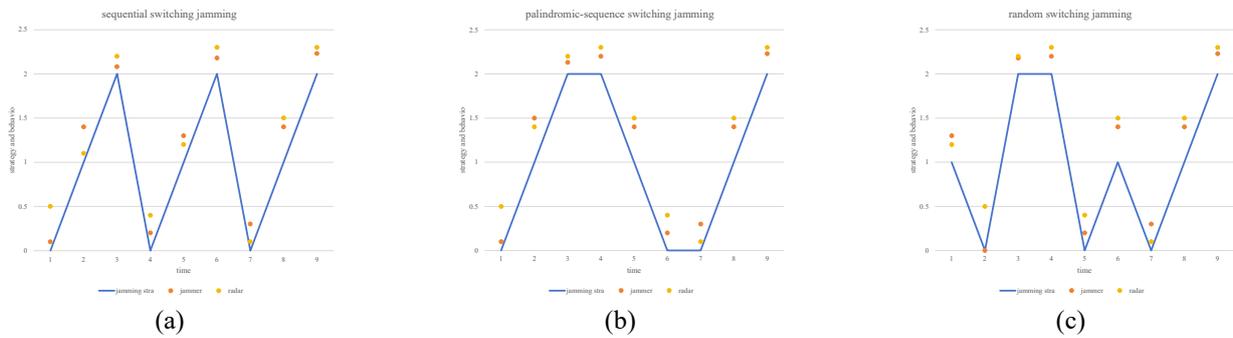

Fig. 14. Selected radar waveforms under different jamming scenarios. (a) Sequential switching jamming. (b) Palindromic-sequence switching jamming. (c) Random switching jamming.

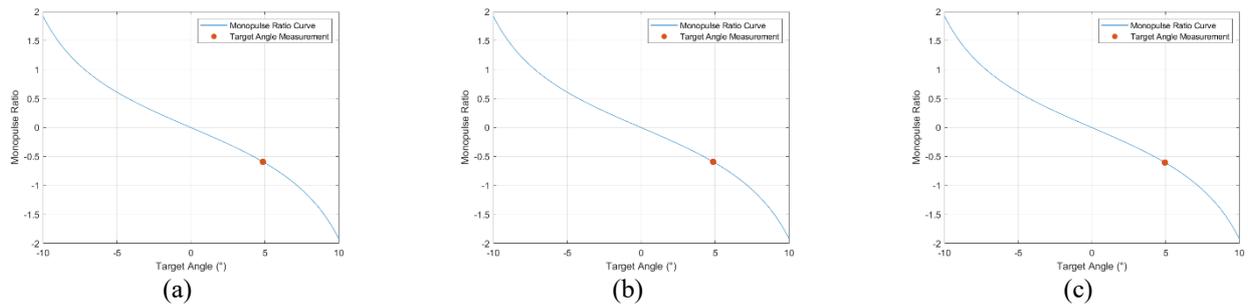

Fig. 15. Monopulse Angle Estimation Results After Anti-Jamming. (a) AJ. (b) RFT. (c) RDFT.

Furthermore, in a scenario where both the target angle is 5° and the jammer angle is 2°, the decision network outputs corresponding suppression measures, successfully mitigating the interference, as shown in Fig. 15.

## V. CONCLUSIONS

This study addresses a fundamental limitation in modern electronic warfare systems: the fragmented design between jamming recognition and anti-jamming decision-making. It severely compromises radar resilience in DRFM-dominated complex electromagnetic environments. To bridge this gap, we propose an integrated anti-jamming framework, which providing a closed-loop anti-jamming control from multiple jamming type recognition to anti-jamming waveform decision-making.

The integration of jamming type classification and intelligent waveform decision-making is realized through active-passive radar fusion, enabling a unified framework for jamming type recognition and countermeasure selection. The recognition part can classify three jamming types, and the decision-making part can make waveform decision to suppress jamming. The proposed method is compared with related methods, including SVM, VGG16, 2D-CNN, and SARSA model. Its performance in the recognition part is evaluated based on OA, Recall, Precision, and F1. The results show that the proposed method demonstrates noticeable improvements across the evaluated metrics compared to the baseline methods. Especially, during the subsequent intelligent waveform decision-making part, it achieves more rapid convergence and maintains high stability. In terms of jamming recognition, the use of multi-modal feature fusion improves recognition accuracy. In the intelligent decision-making module, active-passive radar fusion enhances operational stability and enables faster waveform selection. Furthermore, we constructed an application scenario where the target angle is 5° and the jammer angle is 2°. Subsequently, the echo data after interference suppression were processed using monopulse angle estimation, resulting in improved angle measurement accuracy. The proposed method was applied for interference suppression in this scenario.

In future work, we aim to implement multi-platform cooperative operations by integrating airborne and ground-based radar systems. This will enhance anti-jamming performance, and support more complex combat scenarios.

## REFERENCES

[1] Soumekh M. SAR-ECCM using phase-perturbed LFM chirp signals and DRFM repeat jammer penalization[C]//IEEE International Radar Conference, 2005. IEEE, 2005: 507-512.

[2] Yang B, Li K, Jiu B, et al. An intelligent jamming strategy design method against frequency agility radar[C]//2023 IEEE International Radar Conference (RADAR). IEEE, 2023: 1-6.

[3] Tan M, Wang C, Xue B, et al. A novel deceptive jamming approach against frequency diverse array radar[J]. IEEE Sensors Journal, 2020, 21(6): 8323-8332.

[4] Wang W, Wu J, Pei J, et al. An antideceptive jamming method for multistatic synthetic aperture radar based on collaborative localization and spatial suppression[J]. IEEE Journal of Selected Topics in Applied Earth Observations and Remote Sensing, 2020, 13: 2757-2768.

[5] Neng-Jing L, Yi-Ting Z. A survey of radar ECM and ECCM[J]. IEEE Transactions on Aerospace and Electronic Systems, 1995, 31(3): 1110-1120.

[6] Waters W M, Linde G J. Frequency-agile radar signal processing[J]. IEEE Transactions on aerospace and electronic systems, 1979 (3): 459-464.

[7] Kogon S M, Holder E J, Williams D B. Mainbeam jammer suppression using multipath returns[C]//Conference Record of the Thirty-First

[8] Deng H, Himed B. Target Detection and Interference Mitigation in Future AI-Based Radar Systems[C]//2021 IEEE Radar Conference (RadarConf21). IEEE, 2021: 1-4.

[9] Richter Y, Balal N, Gerasimov J, et al. Deep Learning-Based Radar Processing: Simultaneous Target Classification, Activity Detection, and Range Estimation with Micro-Doppler Radar[C]//2024 IEEE International Conference on Microwaves, Communications, Antennas, Biomedical Engineering and Electronic Systems (COMCAS). IEEE, 2024: 1-3.

[10] Liu Q, Zhang W. Deep learning and recognition of radar jamming based on CNN[C]//2019 12th international symposium on computational intelligence and design (ISCID). IEEE, 2019, 1: 208-212.

[11] Zhou H, Wang L, Guo Z. Recognition of radar compound jamming based on convolutional neural network[J]. IEEE Transactions on Aerospace and Electronic Systems, 2023, 59(6): 7380-7394.

[12] Liu Q, Zhang W. Deep learning and recognition of radar jamming based on CNN[C]//2019 12th international symposium on computational intelligence and design (ISCID). IEEE, 2019, 1: 208-212.

[13] Hu Z, Li H, Tang Z, et al. Radar Signal Recognition Based on Deep Feature Fusion of Multiple Time-frequency Images[C]//2023 International Conference on Wireless Communications and Signal Processing (WCSP). IEEE, 2023: 377-383.

[14] Dong G, Wang Z, Liu H. A Cross-Modality Contrastive Learning Method for Radar Jamming Recognition[J]. IEEE Transactions on Instrumentation and Measurement, 2025.

[15] Shao G, Chen Y, Wei Y. Convolutional neural network-based radar jamming signal classification with sufficient and limited samples[J]. IEEE Access, 2020, 8: 80588-80598.

[16] Zheng L, Li S, Tan B, et al. Rcfusion: Fusing 4-d radar and camera with bird's-eye view features for 3-d object detection[J]. IEEE Transactions on Instrumentation and Measurement, 2023, 72: 1-14.

[17] Lv Q, Quan Y, Feng W, et al. Radar deception jamming recognition based on weighted ensemble CNN with transfer learning[J]. IEEE Transactions on Geoscience and Remote Sensing, 2021, 60: 1-11.

[18] Shi Y, Sagduyu Y E, Erpek T, et al. Adversarial deep learning for cognitive radio security: Jamming attack and defense strategies[C]//2018 IEEE international conference on communications workshops (ICC Workshops). IEEE, 2018: 1-6.

[19] Zhou Y, Shang S, Song X, et al. Intelligent radar jamming recognition in open set environment based on deep learning networks[J]. Remote Sensing, 2022, 14(24): 6220.

[20] Tang Y, Zhao Z, Ye X, et al. Jamming recognition based on AC-VAEGAN[C]//2020 15th IEEE International Conference on Signal Processing (ICSP). IEEE, 2020, 1: 312-315.

[21] Cai Y, Shi K, Song F, et al. Jamming pattern recognition using spectrum waterfall: A deep learning method[C]//2019 IEEE 5th international conference on computer and communications (ICCC). IEEE, 2019: 2113-2117.

[22] Li S, Liao M, Xiong K, et al. Cooperative Frequency Scheduling for Netted Agile Radars via Decentralized Q-Learning[J]. IEEE Transactions on Instrumentation and Measurement, 2025.

[23] Janiar S B, Wang P. Intelligent anti-jamming based on deep reinforcement learning and transfer learning[J]. IEEE Transactions on Vehicular Technology, 2024, 73(6): 8825-8834.

[24] Wei J, Wei Y, Yu L, et al. Radar anti-jamming decision-making method based on DDPG-MADDPG algorithm[J]. Remote Sensing, 2023, 15(16): 4046.

[25] Cheng S, Ling X, Zhu L. Deep Reinforcement Learning-Based Anti-Jamming Approach for Fast Frequency Hopping Systems[J]. IEEE Open Journal of the Communications Society, 2025.

[26] Jiang W, Wang Y, Li Y, et al. An intelligent anti-jamming decision-making method based on deep reinforcement learning for cognitive radar[C]//2023 26th International Conference on Computer Supported Cooperative Work in Design (CSCWD). IEEE, 2023: 1662-1666.

[27] Ali Z, Rezki Z, Sadjadpour H. Deep-Q reinforcement learning for fairness in multiple-access cognitive radio networks[C]//2022 IEEE Wireless Communications and Networking Conference (WCNC). IEEE, 2022: 2023-2028.

[28] Zhu J, Ruan H, Han L, et al. Adaptive Selection of Cognitive Radar Waveforms Based on Improved Deep Q-Learning Network[C]//2023 3rd International Conference on Computer Science, Electronic Information Engineering and Intelligent Control Technology (CEI). IEEE, 2023: 515-522.

[29] Jiang W, Qi Z, Ye Z, et al. Research on cooperative detection technology of networked radar based on data fusion[C]//2021 2nd China International SAR Symposium (CISS). IEEE, 2021: 1-5.

[30] Jiang W, Qi Z, Ye Z, et al. Research on cooperative detection technology of networked radar based on data fusion[C]//2021 2nd China International SAR Symposium (CISS). IEEE, 2021: 1-5.

[31] Liu Y, Research on Anti-Deceptive Jamming Methods for Multistatic Radar Systems, Ph.D. dissertation, Xidian University, Xi'an, China, 2018. (In Chinese).

[32] Zhao S, Research on Cooperative Anti-Deceptive Jamming Methods for Multistatic Radar Systems, Ph.D. dissertation, Xidian University, Xi'an, China, 2016. (In Chinese).

[33] Jung D H, Kim D H, Azim M T, et al. A novel signal processing technique for Ku-band automobile FMCW fully polarimetric SAR system using triangular LFM[J]. IEEE Transactions on Instrumentation and Measurement, 2020, 70: 1-10.

[34] Lu W, Zhang Q. Deconvolutive short-time Fourier transform spectrogram[J]. IEEE Signal Processing Letters, 2009, 16(7): 576-579.

[35] Hang H. Time-frequency DOA estimate algorithm based on SPWVD[C]//2005 IEEE International Symposium on Microwave, Antenna, Propagation and EMC Technologies for Wireless Communications. IEEE, 2005, 2: 1253-1256.

[36] Iandola F, Moskewicz M, Karayev S, et al. Densenet: Implementing efficient convnet descriptor pyramids[J]. arXiv preprint arXiv:1404.1869, 2014.

[37] Liu Z, Wang L, Wen Z, et al. Multilevel scattering center and deep feature fusion learning framework for SAR target recognition[J]. IEEE Transactions on Geoscience and Remote Sensing, 2022, 60: 1-14.

[38] Wu M. Music Emotion Classification Model Based on Multi Feature Image Fusion[C]//2024 First International Conference on Software, Systems and Information Technology (SSITCON). IEEE, 2024: 1-6.

[39] Raza A, Huo H, Fang T. PFAF-Net: Pyramid feature network for multimodal fusion[J]. IEEE Sensors Letters, 2020, 4(12): 1-4.

[40] Tolstikhin I O, Houlsby N, Kolesnikov A, et al. Mlp-mixer: An all-mlp architecture for vision[J]. Advances in neural information processing systems, 2021, 34: 24261-24272.

[41] Wang S, Liu Z, Xie R, et al. Reinforcement learning for compressed-sensing based frequency agile radar in the presence of active interference[J]. Remote Sensing, 2022, 14(4): 968.

[42] LI Xinzhi and DONG Shengbo. Research on efficient reinforcement learning for adaptive frequency-agility radar[J]. Sensors, 2021, 21(23): 7931. doi: 10.3390/s21237931.

[43] Yao D, Liu Z, Li F, et al. Distorted SAR target recognition with virtual SVM and AP-HOG feature[C]//IET International Radar Conference (IET IRC 2020). IET, 2020, 2020: 734-739.

[44] Xu S, Qi H, Zhang Y, et al. Research on Advanced Detection Method of Target Spectrum of Radar Altimeter Based on VGG16-Net[C]//2023 IEEE International Conference on Signal Processing, Communications and Computing (ICSPCC). IEEE, 2023: 1-6.

[45] Shao G, Chen Y, Wei Y. Deep fusion for radar jamming signal classification based on CNN[J]. IEEE Access, 2020, 8: 117236-117244.

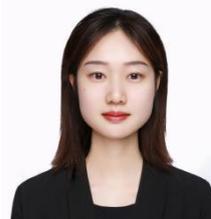

**Huake Wang** received the B.S. degree in electronics engineering, and the Ph.D. degree in signal and information processing from Xidian University, Xi'an, China, in 2015 and 2020, respectively. She was a Visiting Ph.D. Student with the Department of Electrical Engineering, Columbia University, New York, from 2020 to 2021. Currently, she is an Associate Professor with Xidian University. Her research interests include signal processing, new concept radar and intelligent sensing.

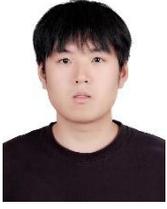
**Xudong Han** received the B.S. degree from Henan University of Science and Technology, Henan, China, in 2023. He is with the Hangzhou Institute of Technology, Xidian University, Hangzhou 311231, China, and also with the Xi'an Key Laboratory of Advanced Remote Sensing, Xi'an 710071, China. His research interests include interference identification and classification, moving target detection, jamming suppression, and multiple-input multiple-output radar signal processing.

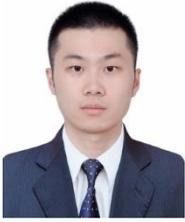
**Bairui Cai** received the bachelor's degree from Hangzhou Dianzi University, Hangzhou, China, in 2023. He is currently pursuing the master's degree at Xidian University, Xi'an, China. He has issued a patent and chaired a graduate Student Innovation fund. His research interests include radar signal processing and deep learning.

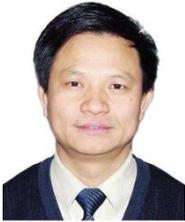
**Guisheng Liao** (Senior Member, IEEE) was born in Guilin, Guangxi, China in 1963. He received the B.S. degree from Guangxi University in mathematics, Guangxi, China, in 1985, and the M.S. degree in computer software and the Ph.D. degree in signal and information processing from Xidian University, Xi'an, China, in 1990, and 1992, respectively.

He is currently a Full Professor with the National Key Laboratory of Radar Signal Processing and served as the 1st Dean with the Hangzhou Institute of Technology, Xidian University since 2021. He has been the Dean with the School of Electronic Engineering, Xidian University from 2013 to 2021. He has been a Senior Visiting Scholar with the Chinese University of Hong Kong from 1999 to 2000. He won the National Science Fund for Distinguished Young Scholars in 2008. His research interests include array signal processing, space-time adaptive processing, radar waveform design, and airborne/space surveillance and warning radar systems.

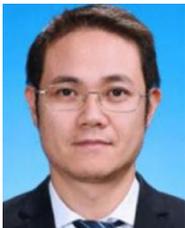
**Yinghui Quan** (Senior Member, IEEE) received the B.S. and Ph.D. degrees in electrical engineering from Xidian University, Xi'an, China, in 2004 and 2012, respectively.

He is currently a Full Professor with the Department of Remote Sensing Science and Technology, School of Information Mechanics and Sensing Engineering, Xidian University. His research interests include agile radar and radar remote sensing.